\documentclass[a4paper,12pt]{article}

\def\be   {\beta}

\def\HH {\hat H}

\def\HH {\hat H}

{\it}
{\it}
{\it}
{\it}
{\it}
{\it}
\def\be {\beta}

\def\de {\delta}

\def\Om {\Omega}

\usepackage{times}
\usepackage{graphicx}
\begin{document}
\title{ Ab-initio calculation of the ${}^6Li$ binding energy  
        with the Hybrid Multideterminant scheme.}
\author{G. Puddu\\
       Dipartimento di Fisica dell'Universita' di Milano,\\
       Via Celoria 16, I-20133 Milano, Italy}
\maketitle
\begin {abstract}
         We perform an ab-initio calculation for the binding energy of
         ${}^6Li$ using the CD-Bonn 2000 NN potential renormalized with the
         Lee-Suzuki method. The many-body approach to the problem is the
         Hybrid Multideterminant method. The results indicate a binding energy 
         of about $31 MeV$, within a few hundreds KeV uncertainty.
         The center of mass diagnostics are also discussed. 

\par\noindent
{\bf{Pacs numbers}}: 21.60.De, $\;\;\;\;$ 21.10.Dr, $\;\;\;\;$ 27.20.+n 
\vfill
\eject
\end{abstract}
\section{ Introduction.}
     A major problem in nuclear physics is the understanding of the structure of 
     nuclei starting from nucleon-nucleon potentials that reproduce the nucleon-nucleon
     scattering data  and the properties of the deuteron. There are nowadays many high accuracy
     nucleon-nucleon potentials that reproduce these data, either phenomenological or based
     on meson exchange theories, such as the Argonne V18 (ref.[1]) and the
     CD-Bonn 2000 (ref.[2]) or, based on chiral perturbation theory, the N3LO (ref.[3])
     NN potential. Accurate predictions at the level of NN potentials are rather important in
     order to elucidate the role of the NNN interaction which are much more difficult to
     use in nuclear structure calculations.
\par
     Once the NN potential is selected, one is left with the many-body problem to evaluate nuclear
     properties. There are two main steps in order to achieve this goal. 
\par\noindent 
     The first step  is to renormalize the NN interaction in order to be able to use small 
      model spaces, and the second one is the many-body problem itself.
     Although for very few nuclei (closed shells) sometimes the bare interaction is used,
     at the price of very large model spaces (ref. [4]), a popular prescription is the Lee-Suzuki method,
     (ref. [5])
     whereby an effective interaction is constructed in a small model space, typically
     using an harmonic oscillator basis, or, as in the case of low momentum interactions,
     a momentum basis (ref. [6] and references in there).
     A limitation of this approach is that many-body interactions are introduced,  
     and usually one keeps only the two-body 
     part of the renormalized interaction (the 2-particle cluster approximation). As a
     consequence the independence of the results from the model space must be checked.
     To further complicate matters, the NN effective
     interaction derived in this way is not unique, especially because of the hermitization
     prescription. Although the
     freedom to hermitize the effective interaction is large, two  prescriptions
     are mostly used, the one of ref. [5] (known as the Okubo hermitization) and the one 
     of ref.[7], mostly used with low momentum interactions. It is worthwhile
     to observe that, at least in principle, this freedom could be used to mimic three-body force
     effects, much in the same spirit it is done with the JISP interactions
     (ref.[8] and references in there).
     This could be very useful especially for low momentum interactions.
\par\noindent 
     The second step is the  solution of the  Schroedinger equation for the nuclei under
     study. Several methods are available. For example
     the no core shell model (NCSM) (ref. [9],[10]), which diagonalizes the Hamiltonian renormalized
     up to a given number of $\hbar\Om$ excitations. Or the coupled
     cluster method (ref.[11] and references in there) whereby the wave function is 
     written as an exponential of one-body+two-body+...
     operators acting on a reference Slater determinant. The first of these methods, although it is
     the most used in ab-initio studies of light nuclei, is limited by  the  large sizes of the Hilbert space.
     The second of these methods, namely the coupled cluster method, is usually applied at or 
     around closed shells. A third type of methods are based on variational schemes, as the VAMPIR method and its 
     variants (ref.[12]),
     the Quantum Monte Carlo method (ref.[13]) and the Hybrid Multideterminant method (HMD) (ref.[14]).
     In this work we shall use
     this last one which is based on the expansion of the nuclear wave function as a sum of
     a large number (as many as the accuracy demands) of symmetry unrestricted Slater determinants (SD)
     with the appropriate angular momentum and parity quantum numbers restored with projectors, the 
     Slater determinants being determined solely by variational requirements.
     This method does not suffer from the limitation of the size of the Hilbert space, it approaches
     more and more the exact ground state wave function as the number of Slater determinants is increased,
     and furthermore it is equally applicable to both closed and open shell nuclei.
     So far it has been applied in a no core fashion using the Argonne v8' NN potential (ref.[14]) and to a 
     phenomenological local potential in order to study shell effects using the bare interaction (ref.[15]).
     It has also been applied to nuclei in the $fp$ region using phenomenological effective interactions
     (ref.[16]),however these systems are relatively easy since the bulk of the energies are of single-particle
     character. 
\par
     In this work we shall apply the HMD method to ${}^6Li$ starting 
     from the accurate CD-Bonn 2000 interaction. This nucleus has been extensively studied within
     the NCSM approach, using both the CDBonn (ref. [17]), the CDBonn 2000 (ref.[18],[19]) and the N3LO interactions
     (ref. [19]). 
     The motivation to perform a calculation for this nucleus using a different many-body method is the following.
     An ab-initio calculation requires the  results to be independent on the size of the model space
     and also  on the value of $\hbar\Om$ of the harmonic oscillator single-particle  
     basis, at least within some range of values. So far the calculations reported in the literature using the 
     Lee-Suzuki renormalization
     prescription show a residual dependence on the value of $\hbar\Om$. Such a dependence is not seen
     using soft potentials such as the low-momentum interaction or the JISP16 interaction (cf. ref. [20]). 
     Eventually such
     a dependence should disappear using larger values of the maximum allowed number of $\hbar\Om$
     excitations ($N_{max}$). The HMD method does not use $\hbar\Om$ excitations, but rather utilizes an
     Hamiltonian in a specified number of major harmonic oscillators shells, which contain a much larger
     (although not all possible) $N_{max}$ excitations. We do obtain a weaker dependence
     on $\hbar\Om$, but the dependence does not disappear at large value $\hbar\Om$. However we obtain
     a much lower value for the ground-state energy, closer to the experimental value.
\par
     The HMD method, in its ab-initio form, can be formulated in two different ways.
     One can construct the effective Hamiltonian directly in the lab frame for a specified number of
     harmonic oscillator major shells (up to $N_s$ total quantum number)  using the standard 
     Talmi-Moshinsky brackets (cf. for example ref.[21]) relating these matrix elements to the renormalized matrix elements 
     in the center of mass frame (HMD-a version).
     In this case the renormalized matrix elements in the center of mass frame up to $N_{cm}=2N_s$
     total harmonic oscillator quantum number in the center of mass frame are needed.
     Differently  one could first construct the matrix elements of the 
     renormalized Hamiltonian using $N_{cm}+1$ harmonic oscillator shells and then transform the Hamiltonian
     to the lab frame using the same number $N_{cm}+1$ of harmonic oscillator shells (HMD-b version). 
     The difference between the HMD-a and the HMD-b version consists in the fact that the HMD-a version
     truncates the Hamiltonian used in the HMD-b version. Conversely a large fraction of the matrix elements
     of the renormalized Hamiltonian used by HMD-b are set to $0$, more precisely all matrix elements 
     of the 
     type $<ab|H_{eff}|cd>$ for which the states $a,b$ or $c,d$ satisfy the relation $ 2 n_a+l_a+2 n_b+l_b>N_{cm}$
     ($n,l$ being the harmonic oscillator quantum numbers).
\par
     The HMD-b version for $A=2$ is exact in the 
     sense that 
     reproduces to very high accuracy the eigenvalues of the bare Hamiltonian, while the HMD-a version
     converges to the exact values only in the limit of a large number of harmonic oscillator shells.
     As a consequence the HMD-a version needs to be validated. For $A=2$ clearly HMD-b is superior, however
     we find that for $A=3$,  HMD-b overbinds and that the HMD-a version is superior even for a smaller number of major
     harmonic oscillator shells. This can be understood by recalling that both versions neglect 3-particle cluster
     contributions to the renormalized interaction and the implication is therefore that HMD-a has smaller
     3-particle cluster effects. In other words, the truncation performed in the HMD-a version effectively takes
     into account at least some of the missing 3-body interaction  induced by an exact renormalization, while
     in the HMD-b version this can be done only by increasing the number of major shells. This is of course
     a useful result, although empirical.
     For ${}^6Li$ we prefer to use the HMD-a version, since also for this nucleus HMD-b strongly overbinds 
     even compared to the experimental binding energy.
\par
     The outline of this paper is the following. In section 2 we discuss the validation of the two versions
     and of the computer programs and in section 3 we discuss the case of ${}^6Li$ and also the center of
     mass diagnostic recently proposed in ref. [22]. We also discuss a calculation for the $3^+$ excited
     state.
\bigskip
\bigskip
\section{ Validation of the method.}
\bigskip
     Both versions of the HMD method start, as in NCSM approach (refs.[9],[10]), from the Hamiltonian
$$
\HH=\sum_{i=1}^A {p_i^2\over 2m }+ \sum_{i<j} V_{ij}= \HH_{int}+ {P_{cm}^2\over 2 mA},
\eqno(1)
$$
      $m$ being the average nucleon mass for the nucleus under consideration, $V$ the nucleon-nucleon potential, 
      $P_{cm}$ is the total momentum and $\HH_{int}$ is the intrinsic Hamiltonian. As in ref. [9], to this Hamiltonian 
      an harmonic potential acting on the center of mass is added, that is 
$$
\HH_{\Om}=\HH_{int}+\HH_{cm}=\HH+{1\over 2} mA\Om^2 R_{c.m.}^2=\sum_{i=1}^A h_i +\sum_{i<j}  V_{ij}^{(A)},
\eqno(2)
$$
      with
$$
 V_{ij}^{(A)}=V_{ij} -{m \Om^2\over 2 A}r_{ij}^2,
\eqno(3)
$$
      and
$$
h_i= {p_i^2\over 2 m} + {1\over 2} m \Om^2 r_i^2.
\eqno(4)
$$
     $\HH_{cm}$ in eq.(2) is the harmonic oscillator Hamiltonian of the center of mass
$$
\HH_{cm} = {P_{cm}^2\over 2 mA} + {1\over 2} mA\Om^2 R_{c.m.}^2.
\eqno(5)
$$

     The Hamiltonian of eq.(2), in which $A$ is  considered as a parameter, is solved for the two-particle systems
     in an harmonic oscillator basis using a large number of major shells (typically $400\div 500$) in all possible
     angular momentum isospin and z-projection of the isospin channels $j s t t_z$ in the intrinsic frame
     of the two-particle system. The number of major shell is taken large enough so that the Hamiltonian can be considered 
     in the "infinite" space (the P+Q space). All integrals are evaluated using typically $2000$ integration points.
     After having done this, the Lee-Suzuki
     (with the Okubo hermitization) renormalization prescription is performed in which the model space is
     restricted to the first $N_{cm}+1$ major harmonic oscillator shells (the P space) of the intrinsic frame (cf. also 
     ref. [23] for a very 
     compact derivation). $N_{cm}$ is taken to be even, as it will clear in the following ($N_{cm}=2N_s$).
     Once the renormalized $A$-dependent Hamiltonian for the two-particle system is obtained, the two-body matrix elements 
     of the effective interaction are extracted and the matrix elements of the intrinsic Hamiltonian of the $A$  particle 
     system (the original nucleus) are evaluated.
\par
     The HMD method can now be branched into two. The two-body matrix elements for the nucleus under consideration
     can be transformed into the lab frame up to $N_s+1$ major shells (HMD-a version), or can be transformed into the lab 
     frame up to $N_{cm}+1$ major shells (HMD-b version). The situation is schematically illustrated in fig. 1.
     In the HMD-b version all matrix elements having one state in the upper right triangle are set to 0.
     One can optionally add to the lab frame Hamiltonian a term $\be (H_{cm}-3/2 \hbar \Om)$ as commonly done.
     The effect of this term due to finite space sizes has been recently analyzed in ref. [22] in order 
     to study
     unphysical couplings between intrinsic modes and center of mass excitations (cf. next section also).
     In both HMD-a and HMD-b versions the resulting Hamiltonian is the input for a variational calculation as done
     in ref. [14]. The variational method in the most recent computer programs is the one discussed in refs. [14],[24].
     The wave function is a linear combination of Slater determinants (without symmetry restrictions) with good
     quantum numbers restored by projectors.

\renewcommand{\baselinestretch}{1}
\begin{figure}
\centering
\includegraphics[width=10.0cm,height=10.0cm,angle=-90]{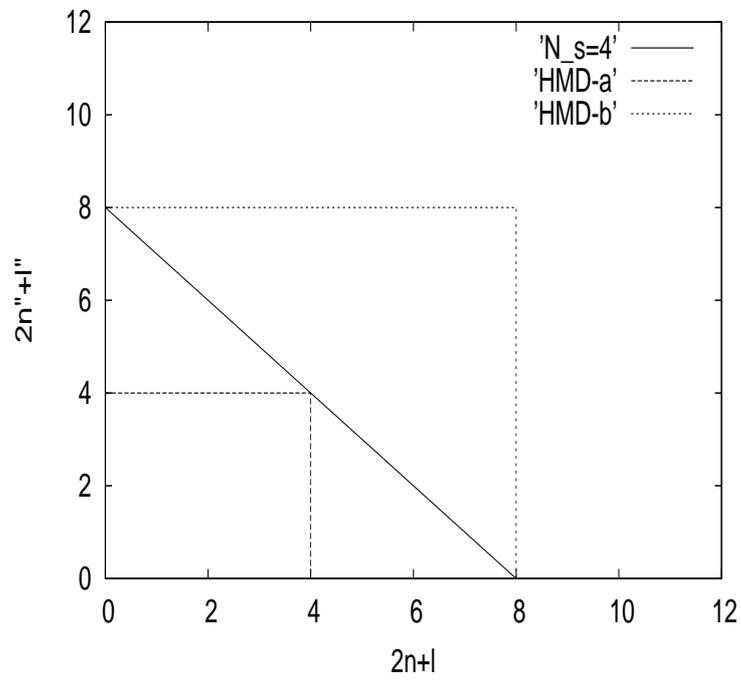}
\caption{Schematic representation of the model spaces used in the HMD-a and HMD-b for $N_s=4$
In the HMD-b all matrix elements in the upper triangle are to 0.}
\end{figure}
\renewcommand{\baselinestretch}{2}
     Needless to say HMD-a is  computationally cheaper than HMD-b. A 5 major shells calculation with 
     HMD-a translates into a 9 major shells calculation with HMD-b, for example.  
     The details of the optimization techniques will discussed in the next section, since they are the same
     utilized for the validation.
     The validation of the whole set of the computer codes is performed first on Deuterium. Actually in this
     (and only in this case) a numerical cancellation in the renormalization step prevents the exact reproduction
     of the "bare" eigenvalues. For all other nuclei, the renormalization step reproduces
     the "bare" eigenvalues belonging to the model space to very high accuracy. For $\hbar\Om = 16 MeV$ with
     $N_{cm}=8$ we obtained the renormalized binding energy of deuterium with an error of $0.26 eV$ using 15 Slater 
     determinants
     (projected to $J_z^{\pi}=1^+$) using the version HMD-b. The situation is different for the HMD-a version
     since not all matrix elements in the intrinsic frame are used. We therefore expect that the variational
     calculation will reproduce the renormalized binding energy only in the limit of large $N_s$. We performed
     some tests for $\hbar\Om= 12 MeV$. For $N_s=4$ the difference between the binding energy obtained by the
     variational calculation and the exact value is $\de=0.041 MeV$, for $N_s=5$, we obtained $\de=0.026 MeV$ and
     for $N_s=7$ (excluding all states with $l=7$) we obtained $\de=0.012 MeV$. This test validates both versions
     of the methods.
\par
      We performed also some tests for ${}^3H$ and ${}^4He$. For ${}^3H$ the binding energy obtained
     with the Faddeev equation method (ref. [25]) 
     using the CD-Bonn 2000 interaction, is $-7.998 MeV$. 
     In this case both versions can reach the exact value only in the limit of large $N_s$ (or $N_{cm}=2N_s$).
     For the HMD-a version and  $\hbar\Om=16 MeV$, we obtained a ground state energy (in MeV) of 
     $ -8.29, -8.30, -8.14, -8.03$ for $ N_s=3, N_s=4, N_s=5 $ and $N_s=6$ respectively. For low $N_s$, about $35\div 50$
     Slater determinants (with the $J_z^{\pi}$ projector) are needed to converge. For large $N_s$ the number of Slater 
     determinants is larger.
     For $\hbar\Om=18 MeV$, the ground-state energy in MeV is $-8.183$, $-8.176$, $-8.125$ and $-7.961$ for
     $ N_s=3, N_s=4, N_s=5 $ and $N_s=6$ respectively. As before, the calculations for large model space are
     more involved and a large number of Slater determinants is necessary.
     We estimate a possible further decrease in the energy of few tens of $KeV$.
     For larger values of $\hbar\Om$ the calculation  becomes increasingly more difficult for large model 
     space.
     For $\hbar\Om=20 MeV$ we obtained for the ground-state energy (in MeV) $-8.023$, $-8.044$, $-7.914$ for 
     $ N_s=3, N_s=4, N_s=5 $
     respectively. The wave functions obtained with the HMD-a version can serve as a variational input for the 
     HMD-b version with $N_{cm}=2N_s$. For this version we performed only few calculations since the model spaces
     are very large and the omission  of large $l$ values of the single-particle orbits is necessary. As an example
     for $\hbar\Om=16 MeV$ and $N_{cm}=6$ omitting all single-particle states having $l$ values larger than $4$
     and using only $ 15$ Slater determinants we obtained a ground-state energy of $-8.843 MeV$. The inclusion of larger 
     $l$-values and the increase of the number of Slater determinants will necessarily lower the energy. This value 
     should  be compared
     with the value obtained with the HMD-a version which is much closer to the exact Faddeev result.
\par
     The only source of discrepancy between the Faddeev result and the HMD-b result comes from the missing 3-particle
     cluster contributions. The conclusion that we can draw is that the missing 3-particle cluster contributions
     are strongly repulsive. The effect of such contributions is much smaller in the HMD-a version. One expects that
     in order to suppress such contributions in the HMD-b implementation one has to increase the number of major shells.
      For $\hbar\Om=18 MeV$ and $N_{cm}=8$ we obtained a ground-state energy of $-8.574 MeV$, in this case we excluded
     from the calculation all $l>6$ values. The inclusion of these states will necessarily decrease the energy.
     The conclusion we can draw form these calculations is that the HMD-b version, although in principle more rigorous,
     strongly overbinds since it misses 3-particle cluster contributions, which seem less relevant in the
     HMD-a version. We performed a calculation also
     for ${}^6Li$ using the HMD-b version, but even without full convergence to a large number of Slater determinants
     we obtained strong overbinding.
     As done in all past calculations with the HMD method, we therefore use only the HMD-a implementation, It is  
     inaccurate only for the 2-particle system, but that is hardly relevant for many-body problems.
\par
     Using the HMD-a approach we performed a calculation for the binding energy of ${}^4He$. We considered
     a reasonable value of the harmonic oscillator frequency, $\hbar\Om=20 MeV$, rather than a full set
     of frequencies, and took $N_s=3,4,5,6,7$. The ground-state energies are (in MeV)
     $E=-29.259, -28.504, -27.603, -26.938$ and $ -26.354$ for $N_s=3,4,5,6,7$ respectively. The calculations
     become increasingly time consuming for large values of $N_s$. In the case of $N_s=6$ we
     built $150$ Slater determinants using the partial $J_z^{\pi}=0^+$ projector and later reprojecting
     the energies using the full angular momentum projector. For $N_s=7$ we took only $100$ Slater determinants.
     The uncertainty in the calculation are about $100 KeV$ or less and $140 KeV$ for $N_s=7$. The
     ncsm result from ref. [27] is $-26.16 MeV$, indicating that for $\hbar\Om=20 MeV$ a larger number of major
     shells are necessary for good accuracy.
\bigskip
\bigskip
\section{ ${}^6Li$.}
\bigskip
\bigskip
     The nucleus ${}^6Li$ with the CDBonn-2000 interaction has been studied in the past in the framework
     of the NCSM method (ref. [18],[19]). The ground-state energy obtained  with 
     this method is $-29.07 MeV$ (the experimental value from ref. [26] is $-31.994 MeV$). The ab-initio approach imposes 
at least for 
     some $\hbar\Om$ interval constancy of the energies as the model space sizes are increased, and as
     $\hbar\Om$  is varied.
     We performed several calculations for this nucleus. The most relevant ones are the ones concerning
     the intrinsic energy. Most often a center of mass term of the type
     $\be (\hat H_{cm} -3\hbar\Om /2)$,
      where $\hat H_{cm}$ is the center of mass harmonic oscillator Hamiltonian,
     is added to the intrinsic Hamiltonian. The effects of the addition of such a term has been recently
     scrutinized in ref.[22] and the unphysical coupling between intrinsic and center of mass Hamiltonian
     caused by the finite size of the model space, has been assessed. It was found in ref.[22] that this unphysical
     coupling using model space defined by a specified number of major shells can decrease the 
     binding  energy in an appreciable way. Here the calculations with the HMD-a method are performed
     using the intrinsic Hamiltonian. The effect of the addition of the center of mass Hamiltonian 
     will be analyzed at the end of the section. The HMD-a calculations proceed in two phases. In the 
     first phase a large number of Slater determinants, typically $100\div 400$
     is generated using only a partial angular momentum and parity projector to good $J_z^{\pi}=1^+$.
     In the second phase this set is reprojected using the full angular momentum and parity projector
     $J^{\pi}=1^+$.
     At least for this nucleus and for this interaction, we find this optimization technique 
     computationally more efficient than
     performing from the beginning the variational calculations with the full angular momentum and 
     parity projector. 
\par\noindent
    The first phase is a combination of two steps. We first  increase the number of Slater 
     determinants (SD) $N_D$ and optimize the last added SD using the steepest descent method, much in the same
     way it has been done in ref. [14]. In the second step we vary anew all SD's one at a time using the 
     quasi-newtonian rank-3 update of ref. [24]. This second step is repeated several times until the
     energy decrease is less than a specified amount. Afterwards, the addition step is repeated.
     We test the accuracy of the final wave function by plotting the energy vs $1/N_D$. As it will be shown,
     for large $N_D$ in many cases the energy is linear in $1/N_D$.
\par
     The total number of SD necessary to obtain a reasonable convergence
     varies depending  on the model space (typically $N_D$ increases as $N_s$ is increased and
     the variational problem becomes harder as $\hbar\Om$ is increased). It does not seem that $N_D$
     depends in any obvious way from the sizes of the Hilbert space which can become very large as $N_s$ is
     increased. Actually one the main reasons for using methods such as the HMD, is that the calculations can
     be performed even for very large size of the Hilbert space. However feasibility does not necessarily
     imply accuracy, as the value of $N_D$  necessary to reach a given accuracy could
     depend on the size of the Hilbert space. We performed a test using a set of 400 SD, for the same interaction,
     obtained as a part of another calculation for ${}^{12}C$  with $N_s=3$ (not discussed in this work),
      $\hbar\Om=15 MeV$ and $\beta=0.5$. A 
     reprojection was performed as explained above. 
     For ${}^6Li$ typical size
     of the Hilbert space range from about $10^5$ for $N_s=2$ to about $10^8$ for $N_s=4$, while for ${}^{12}C$ at $N_s=3$
     the size of the Hilbert space is about $10^{12}$.
\renewcommand{\baselinestretch}{1}
\begin{figure}
\centering
\includegraphics[width=10.0cm,height=10.0cm,angle=-90]{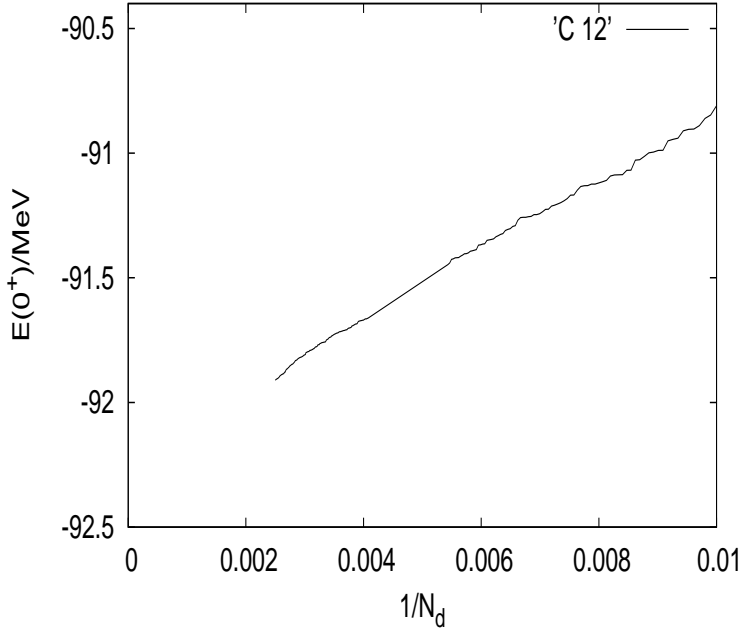}
\caption{Ground-state energy of ${}^{12}C$ as a function of the inverse of the number of Slater determinants 
 for $\hbar\Om=15 MeV$, $N_s=3$ and $\beta=0.5$.}
\end{figure}
\renewcommand{\baselinestretch}{2}
      The calculated value for the ground-state energy of ${}^{12}C$ is $-91.91 MeV$
     (to be compared with the experimental value of $-92.162 MeV$). In fig. 2 we show the behaviour of $E(1/N_D)$
     for large $N_D$. A linear extrapolation shows that a plausible final  energy 
      of $-92.3 MeV$.  A similar behavior is also seen for ${}^6Li$. For comparison
     in fig. 3 we show the behavior of $E(1/N_D)$ for ${}^6Li$ at $\hbar\Om=15 MeV$ and $N_s=4$ with $\beta=0$.    
     Since there is increase of several orders of magnitude in the size of the Hilbert space from ${}^6Li$ to
     ${}^{12}C$ it is reasonable to conclude that if there a dependence of $N_D$ on the size of the Hilbert space,
     such a dependence is very mild. The behavior of the energy as a function of $1/N_D$ can change for different $N_s$
     in the vicinity of the origin. Sometimes the energy behaves as a higher power of $1/N_D$ especially for small $N_s$.
     We performed calculations for ${}^6Li$ for $\hbar\Om=10 MeV,\;\;\;12.5 MeV,\;\;\;15 MeV,\;\;\;17.5 MeV,\;\;\;20 MeV$.
\renewcommand{\baselinestretch}{1}
\begin{figure}
\centering
\includegraphics[width=10.0cm,height=10.0cm,angle=-90]{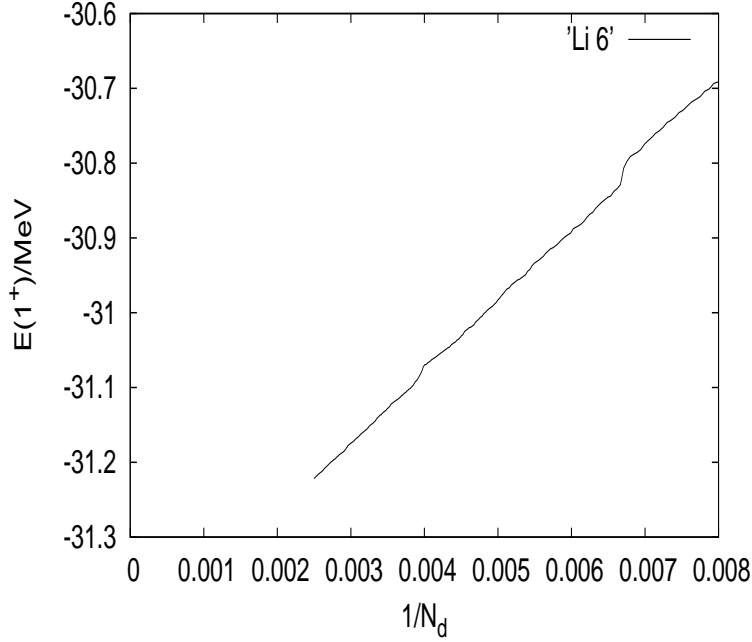}
\caption{Ground-state energy of ${}^{6}Li$ as a function of the inverse of the number of Slater determinants 
 for $\hbar\Om=15 MeV$, $N_s=4$ and $\beta=0$.}
\end{figure}
\renewcommand{\baselinestretch}{2}
     The results are presented in the table. The results for $N_s=2$ and $N_s=3$ are well converged.
      For $N_s=2$ good convergence is reached using $150$ SD's (however for $\hbar\Om=20 MeV$ we had to use $180$ SD's.
     For $N_s=3$ we used $400$ SD's ($450$ for $\hbar\Om=20 MeV$) and also for $N_s=4$. The results for $N_s=5$
     should be considered as partial ones (we used a set of 300 or less Slater determinants). In fact the computational 
    cost of the variational calculation depends
     mostly on the size of the single-particle space. The dependence on the particle number is rather mild.
\renewcommand{\baselinestretch}{1}
\begin{table}
        \begin{tabular}{| c | c | c | c | c |}
                \hline
$ \hbar\Om(MeV)$ &   $ N_s=2      $   & $   N_s=3     $ &   $   N_s=4   $  & $  N_s=5     $  \\
\hline
$  10.0        $   & $  -28.712   $   & $  -28.940    $   & $  -30.14   $   & $  -30.65** $  \\
$  12.5        $   & $  -30.707   $   & $  -30.558    $   & $  -31.18   $   & $  -30.99** $  \\
$  15.0        $   & $  -31.525   $   & $  -31.140    $   & $  -31.22   $   & $   -       $  \\
$  17.5        $   & $  -31.381   $   & $  -30.843    $   & $  -30.57   $   & $   -       $  \\
$  20.0        $   & $  -30.455   $   & $  -30.097    $   & $  -29.55*  $   & $   -       $  \\
\hline
\end{tabular}
 \caption { Ground-state energies for ${}^6Li$ for different values of $ \hbar\Om(MeV) $
  and different model spaces $N_s$. Energies are in MeV.
  $*$  Result not fully converged.
  $**$ Only 300 SD were used. For $N_s=4$, 400 SD were employed.}
\end{table}
\renewcommand{\baselinestretch}{2}
\par
      The calculations for ${}^6Li$ were performed without the center of mass Hamiltonian 
      $\HH'=\beta(\HH_{cm}-3/2\hbar\Om)$, 
      i.e. $\beta=0$.
      In ref. [22], The problem of the effect of the addition of $\HH'$ was studied. The main 
      point in
      ref. [22] was that the addition of this term can significantly change the evaluation of the intrinsic 
      energies.
      To be more precise, In a finite space, the eigenstates $|\psi(\beta)>$ of $\HH_{int}+\HH'$ are not a
      product of intrinsic eigenstates and center of mass eigenstates. Thus the intrinsic energies,
      defined as $ E(\beta)=<\psi(\beta)| \HH_{int} |\psi(\beta)>$ acquire  a $\beta$ dependence. These considerations
      do not apply to the calculations for ${}^6Li$ discussed in this work for the following reason.
      Our wave-functions are obtained by minimizing the energy expectation value of  $\HH_{int}$. Therefore,
      since the wave-functions contain $3 A$ space variables, it must factorize into a product of the intrinsic 
      eigenstate and a function (not necessarily an eigenstate) of the center of mass coordinates. The only requirement
      is that good convergence must be reached.
\par\noindent
      One can verify, however, the amount of contamination caused by $\HH'$
      to the intrinsic energies by first minimizing the expectation values of 
      $\HH_{int}+\HH'$ in order to obtain the wave functions $|\psi(\beta)>$,  by evaluating the 
expectation values 
      of $\HH_{int}$ with  $|\psi(\beta)>$ and then by comparing the energies obtained in this way with the real intrinsic
      energies.
      Actually, it is easy to do 
      slightly better than this because of the structure of the HMD ansatz for the wave-functions which are a linear 
      combination of Slater determinants (intrinsic states). The coefficients of this linear combination can easily be
      determined anew in such a way to minimize the intrinsic energy without a re-variation of the intrinsic states.
      As an example we consider $N_s=2$ and $\hbar\Om=15 MeV$ and $\beta=1$. The ground-state energy of
      $\HH_{int}+\HH'$ is $-30.354 MeV$ (obtained with 150 SD's), while the intrinsic energy obtained using
      this eigenstate of $\HH_{int}+\HH'$ is $-31.066 MeV$ (the coefficients of each SD was redetermined).
      This value should be compared with the value given in  the table of $-31.525 MeV$. The discrepancy,
      almost $500 KeV$, is appreciable.
      For this case, i.e. $N_s=2$ $\hbar\Om=15 MeV$ we show in fig. 4 the behavior of $E(\beta)$ as a function of 
      $\beta$.
\par
\renewcommand{\baselinestretch}{1}
\begin{figure}
\centering   
\includegraphics[width=10.0cm,height=10.0cm,angle=-90]{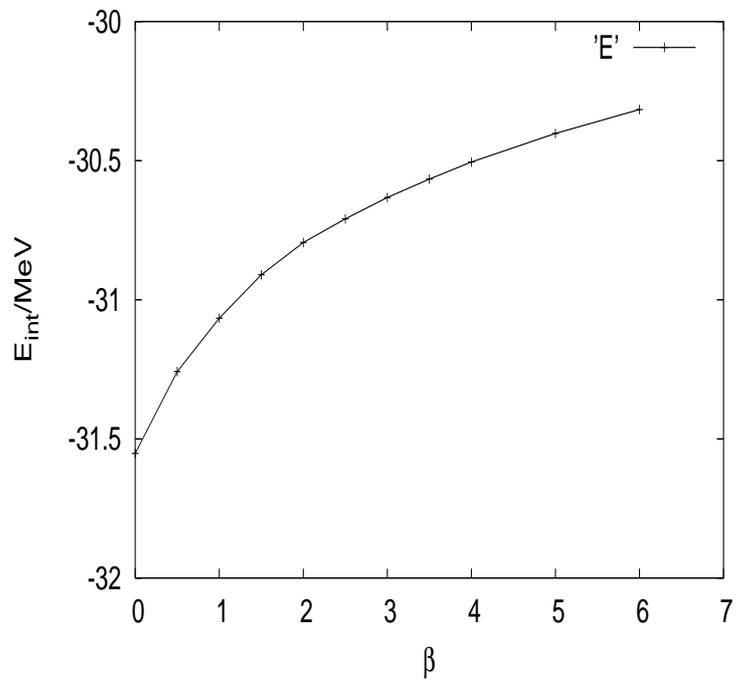}
\caption{$E_{int}$ for ${}^6Li$ for  several $\beta$ values with $N_s=2$ and
 for $\hbar\Om=15 MeV$}
\end{figure}
\renewcommand{\baselinestretch}{2}
\par
\renewcommand{\baselinestretch}{1}
\begin{figure}
\centering   
\includegraphics[width=10.0cm,height=10.0cm,angle=-90]{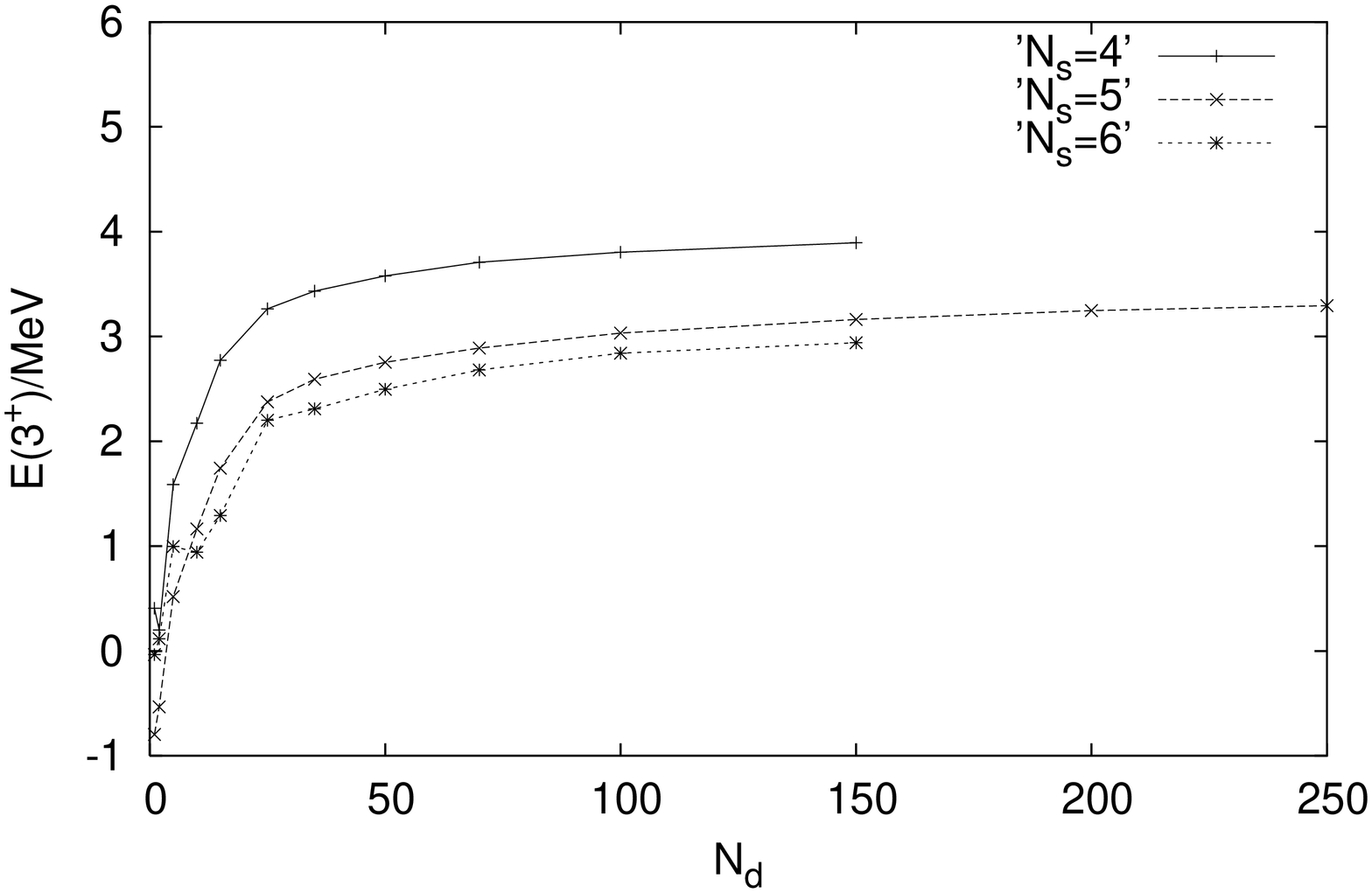}
\caption{$E(3^+)$ for ${}^6Li$ for   $N_s=4,5,6$ and
 for $\hbar\Om=12.5 MeV$}
\end{figure}
\renewcommand{\baselinestretch}{2}
\par
      We also performed a calculation for the excitation energy of the first $3^+$ state,
      by re-evaluating the $J_z^{\pi}=1^+$ and $J_z^{\pi}=3^+$ states using exactly
      the same numerical steps (this is necesssary since both states contain some error
      compared to the values for $N_d=\infty$ and these errors cancel out provided the
      same numerical steps are taken for both states). Only the $J_z^{\pi}$ projector has been
      used. In fig. 5 we show the excitation energy for the $3^+$ state as a function of the
      number of Slater determinants for $N_s=4,5,6$. The value obtained for $N_s=6$ is 
      $2.9 MeV$ higher than the experimental value of $2.18 MeV$, but consistent with the
      ncsm value of $2.86 MeV$.
\par
      In conclusion, we have performed an ab-initio calculation of the binding energy of ${}^6Li$ with the
      Hybrid Multideterminant method in a form that has small 3-particle cluster contributions. The  
      evaluated binding energy is about $31 MeV$ with an uncertainty of few hundreds KeV. This estimate for
      the CD-Bonn 2000 interaction is closer to the experimental value than previously thought. 
\vfill
\eject

\vfill
\eject
\end{document}